\RequirePackage{fix-cm}
\documentclass[twocolumn]{svjour3}          
\smartqed  
\usepackage{graphicx}
\begin{document}

\title{NNLO QCD predictions of the electron charge asymmetry for the inclusive $pp \rightarrow W+X$ production in the forward region at 13 TeV and 14 TeV
}


\author{Kadir Ocalan}


\institute{K. Ocalan \at
              Necmettin Erbakan University, Faculty of Aviation and Space Sciences, Konya, Turkey  \\
              ORCID ID: http://orcid.org/0000-0002-8419-1400 \\
              \email{kadir.ocalan@erbakan.edu.tr}           
}

\date{Received: date / Accepted: date}

\maketitle

\begin{abstract}
This paper presents perturbative QCD predictions of the electron charge asymmetry for the inclusive $W^{\pm}+X \rightarrow e^{\pm} \nu +X$ production in proton-proton (pp) collisions. Perturbative QCD calculations are performed at next-to-next-to-leading order (NNLO) accuracy by using different parton distribution function (PDF) models at 8 TeV, 13 TeV, and 14 TeV center-of-mass energies of the CERN LHC pp collisions. The NNLO calculations are performed for the electrons to have transverse momentum more than 20 GeV in the forward electron pseudorapidity region $2.0 \leq \eta_{e} \leq 4.25$. The NNLO predictions are first compared at 8 TeV with the measurement of the LHCb experiment at the LHC for the W$^{+}$/W$^{-}$ cross section ratio and charge asymmetry distributions. The 8 TeV predictions using NNPDF3.1, CT14, and MMHT2014 PDF sets are reported to be in good agreement with the LHCb data for the entire $\eta_{e}$ region justifying the extension of the calculations to 13 TeV and 14 TeV energies. The charge asymmetry predictions at NNLO accuracy are also reported in the forward $\eta_{e}$ bins at 13 TeV and 14 TeV and compared among NNPDF3.1, CT14, and MMHT2014 PDF sets. Overall the predicted W$^{\pm}$ differential cross section and charge asymmetry distributions based on different PDF sets are found to be consistent with each other for the entire $\eta_{e}$ region. The charge asymmetry distributions are shown to be more sensitive to discriminate among different PDF models in the 14 TeV predictions.     
\end{abstract}

\section{Introduction}
\label{intro}
Production of W and Z bosons has been an important benchmark process to measure at past colliders and in proton-proton (pp) collisions at the CERN Large Hadron Collider (LHC). Precise measurements of their production cross sections provide important tests for the quantum chromodynamic (QCD) and electroweak (EW) sectors of the Standard Model (SM) and valuable inputs to constrain parton distribution functions (PDFs) in the proton. Their measurements enable improvements in background modeling for several rarer SM and beyond the SM processes as W and Z boson productions constitute a major background for those processes. In addition, W and Z boson processes are extensively used for calibrating detector response to improve reconstruction of leptons, jet, and missing energy signature with better performances. W and Z bosons are produced in abundance in their leptonic decay channels with larger cross sections and clean experimental signatures in pp collisions. 

Experimentally, W bosons are reconstructed through their leptonic decay channels $pp \rightarrow W^{\pm} \rightarrow l^{\pm}\nu$. A typical W boson event is characterized by one isolated charged lepton with high transverse momentum $p_{T}$ and large missing transverse energy due to the neutrino. W bosons are produced primarily through the annihilation of a valence quark from one of the colliding protons with a sea antiquark from the other: $u\bar{d}\rightarrow W^{+}$, $d\bar{u}\rightarrow W^{-}$. W$^{+}$ bosons are produced more often than W$^{-}$ bosons due to the presence of two valence $u$ quarks in the proton. This leads to a production asymmetry between W$^{+}$ and W$^{-}$ bosons, which is referred to as W boson charge asymmetry and is usually defined in terms of cross sections $\sigma(W^{+})$ and $\sigma(W^{-})$ differential in the W boson rapidity $y_{W}$ as 

\begin{equation}
\label{eqn:1}   
A(y_{W})=\frac{d\sigma(W^{+})/dy_{W}-d\sigma(W^{-})/dy_{W}}{d\sigma(W^{+})/dy_{W}+d\sigma(W^{-})/dy_{W}}.
\end{equation}

The important capability of this charge asymmetry variable $A(y_{W})$ is to discriminate between PDF models as $y_{W}$ is strongly correlated with the initial-state parton momentum fractions $x$ (Bjorken-$x$ values). However, the neutrino from the W boson decay leaves the detector unobserved and its longitudinal momentum cannot be measured directly. For this reason, the full momentum of the W boson and its rapidity $y_{W}$ cannot be directly reconstructed. In spite of this experimental complication, the same information can be accessed by measuring the charge asymmetry from the W boson decay products. A commonly used approach is to measure the charge asymmetry as a function of the decay lepton (electron\footnote{\label{myfootnote}"Electron" refers to both $e^{+}$ and $e^{-}$ generically throughout the entire paper.}, in this case) pseudorapidity\footnote{\label{myfootnote}Pseudorapidity is defined as $\eta = -ln[(\theta / 2)]$, where $\theta$ is a polar angle relative to the beam axis.} $\eta_{e}$, which is strongly correlated with the $y_{W}$, defined as

\begin{equation}
\label{eqn:2}   
A_{e}=\frac{d\sigma(W^{+}\rightarrow e^{+}\nu)/d\eta_{e}-d\sigma(W^{-}\rightarrow e^{-}\nu)/d\eta_{e}}{d\sigma(W^{+}\rightarrow e^{+}\nu)/d\eta_{e}+d\sigma(W^{-}\rightarrow e^{-}\nu)/d\eta_{e}}.
\end{equation}

The electron charge asymmetry $A_{e}$ variable provides significant constraints on the ratio of $u$ and $d$ quark distributions in the proton as a function of Bjorken-$x$ values of the partons. It can be used to discriminate among PDF models that predict different shapes of valence and sea quark distributions. In addition, this variable can be measured with high accuracy as because of many sources of systematic uncertainties cancel in the ratio, and therefore offers a unique opportunity for precision tests of the SM physics.
 
The W boson production asymmetry was previously measured in p$\rm{\bar{p}}$ collisions by the CDF and D0 Collaborations at the Tevatron~\cite{Abe:1998rv,Abazov:2007pm,Abazov:2008qv,Aaltonen:2009ta,Abazov:2013rja}. W boson lepton charge asymmetry measurements at the LHC were performed in the central lepton pseudorapidity region $|\eta_{l}|\leq$ 2.5 by the ATLAS and CMS Collaborations at different center-of-mass energies up to 8 TeV~\cite{Aad:2011dm,Chatrchyan:2012xt,Chatrchyan:2013mza,Khachatryan:2016pev,Aaboud:2016btc,Aad:2019bdc,Aaboud:2018nic,Aad:2019rou}. The LHCb Collaboration at the LHC has also reported W boson lepton charge asymmetry measurements at 7 TeV~\cite{Aaij:2012vn,Aaij:2014wba} and 8 TeV~\cite{Aaij:2015zlq,Aaij:2016qqz} in the forward lepton pseudorapidity region 2.0 $\leq \eta_{l} \leq$ 4.5 that go beyond the ATLAS and CMS acceptance. In all these complementary results, measurements are compared with the theoretical predictions up to next-to-next-to-leading order (NNLO) accuracy in perturbative QCD convolved with different PDF models. NNLO predictions were obtained either by using $\emph{FEWZ}$~\cite{Li:2012wna} or $\emph{DYNNLO}$~\cite{Catani:2009sm} program for the theoretical comparisons in these measurements. Among all the experimental results available, the measurements of the differential cross sections and thus production charge asymmetries are of particular importance in the forward region of the detector acceptance. The PDFs exhibit significantly large uncertainties at very low and large $x$ values of the interacting leptons, where $x$ values depend also on the acceptance defined by means of the lepton $\eta_{l}$. Therefore, measurements and theoretical predictions of the W boson production asymmetry in the forward detector acceptance 2.0 $\leq \eta_{l} \leq$ 4.5 offer a unique situation to provide valuable inputs on determining accurate PDFs at small and large $x$ values between $10^{-4}\leq x \leq 10^{-1}$.    

In this work we present precise predictions of the electron charge asymmetry between the processes $pp \rightarrow W^{+}+X \rightarrow e^{+}\nu_{e}+X$ and $pp \rightarrow W^{-}+X \rightarrow e^{-}\bar{\nu_{e}}+X$ in the forward $\eta_{e}$ region 2.0 $\leq \eta_{e} \leq$ 4.25. The calculations are performed at NNLO accuracy in the perturbative QCD expansion by using different PDF models at 8 TeV, 13 TeV, and 14 TeV LHC pp collision energies. The predicted W$^{+}$/W$^{-}$ cross section ratio and charge asymmetry distributions as a function of the $\eta_{e}$ are compared with the 8 TeV measurement by the LHCb Collaboration~\cite{Aaij:2016qqz}. The 8 TeV predictions are validated with the data in the fiducial phase space of the LHCb measurement. Further, the NNLO calculations are extended to the 13 TeV and 14 TeV LHC pp energies for the precise predictions of the charge asymmetry in bins of the $\eta_{e}$. The W$^{+}$ and W$^{-}$ boson differential cross distributions along with their W$^{+}$/W$^{-}$ ratios and the charge asymmetry distributions are predicted and compared among different PDF sets. The charge asymmetry results are then reported and compared among different PDF sets. The results reported in this paper represent the first phenomenological study by means of the NNLO predictions for the electron charge asymmetry in the forward $\eta_{e}$ region 2.0 $\leq \eta_{e} \leq$ 4.25 of 13 TeV and 14 TeV pp collisions.       

\section{Methodology}
\label{meth}
\subsection{Computational setup}
\label{comp}
The calculations of the differential cross sections and charge asymmetries are performed by using the \emph{MATRIX} (v1.0.3) computational framework~\cite{Grazzini:2017mhc,Catani:2009sm}. The framework enables calculations of differential cross sections in the form of binned distributions up to NNLO accuracy in perturbative QCD. The so-called transverse momentum \emph{$q_{T}$}-subtraction method~\cite{Catani:2007vq,Catani:2012qa} is employed within the \emph{MATRIX} computations for the cancellations of infrared divergences that arise at the intermediate stages of the calculations. These divergences are regulated by introducing a fixed cut-off value $r_{cut}=$ 0.0015 (0.15\%) for the residual dependence parameter which is defined as $r=p_{T}/m$ in terms of the $p_{T}$ distribution and invariant mass $m$ for a system of colorless particles. Moreover, all the spin- and color-correlated tree-level and one-loop scattering amplitudes are acquired with the use of the \emph{OpenLoops} tool~\cite{Matsuura:1988sm,Cascioli:2011va,Denner:2016kdg} along with the computations. Further to perform theoretical calculations of the differential cross sections in pp collisions, knowledge of the PDFs is required. The \emph{LHAPDF 6.2.0} framework~\cite{Buckley:2014ana} is used for the evaluation of PDFs from data files in the computations. Different PDF sets are used in the calculations, all based on a constant strong coupling $\alpha_{s}=$ 0.118. The \emph{NNPDF3.1}~\cite{Ball:2014uwa}, \emph{CT14}~\cite{Dulat:2015mca}, and \emph{MMHT2014}~\cite{Harland-Lang:2014zoa} PDF sets are used each at NNLO accuracy. To this end, all predicted results are obtained by treating leptons and light quarks massless in the Fermi constant $G_{F}$ input scheme of the computational setup. The default setup of the \emph{MATRIX} framework is used for the input scheme which encompasses the choices for the relevant SM parameters that are based on $G_{F}$ = 1.16639 x 10$^{-5}$ GeV$^{-2}$ value. 

\subsection{Fiducial acceptance}
\label{fid}
The 8 TeV LHCb measurement that was performed in pp collisions at 8 TeV~\cite{Aaij:2016qqz} is chosen to be the reference for the justification of the calculations of this paper. This reference measurement focuses on the differential cross sections and charge asymmetries in the electron decay channel of the W$^{\pm}$ bosons in the forward detector acceptance as well. The fiducial phase space requirements in the calculations are employed to be in line with the reference LHCb measurement. The electrons are required to have transverse momentum $p_{T}>$ 20 GeV and to lie in the forward pseudorapidity region of $2.0 \leq \eta_{e} \leq 4.25$. There is no need to impose any requirements for the final state hadronic jets as the calculations are considered for the inclusive W boson production. The fiducial acceptance is also relaxed to have no explicit requirement for the missing transverse momentum of the final state (anti)neutrino. Furthermore, no requirement for the W boson transverse mass is imposed unlike to the ATLAS and CMS W boson charge asymmetry measurements performed in the central $\eta_{l}$ detector coverage.           

\subsection{Theoretical uncertainties}
\label{theo}
Cross section calculations in perturbative QCD acquire dependence on the renormalization $\mu_{R}$ and factorization $\mu_{F}$ scales and the numerical results will depend on the choice of these scales. In this paper, the central value for the scales is chosen to be the W boson mass $\mu_{R}$ = $\mu_{F}$ = $m(W)$ = 80.385 GeV in the usual way. The theoretical uncertainties due to the choice of the scales, which are simply called as scale uncertainties, referring to missing higher-order contributions in the calculations are estimated by varying independently the $\mu_{R}$ and $\mu_{F}$ up and down by a factor of 2 around the central value. All possible combinations are considered in the variations while imposing the constraint $0.5 \leq \mu_{R}/ \mu_{F} \leq 2.0$. In the cross section ratio and charge asymmetry calculations, this constraint is generalized to an uncorrelated scale variations while restricting to $0.5 \leq \mu/ \mu^{'} \leq 2.0$ between all pairs of scales. Since the scale uncertainties arise from the missing terms in the calculations, the higher the calculated perturbative order, the smaller the theoretical uncertainties associated with the renormalization and factorization procedure will be. Moreover, PDF and $\alpha_{s}$ uncertainties are also considered. The PDF uncertainties for NNPDF3.1, CT14, and MMHT2014 PDF sets are estimated by using the prescription of the PDF4LHC working group~\cite{Butterworth:2015oua,Buckley:2014ana}. The $\alpha_{s}$ uncertainty is estimated by varying the $\alpha_{s}$ value by $\pm0.001$ around 0.118. Estimated theoretical uncertainties for the 8 TeV inclusive cross sections of the W$^{+}$ and W$^{-}$ boson processes at NNLO accuracy are tabulated in Table~\ref{tab:1} as an example to compare relative size of each type of theoretical uncertainty considered in this paper. Total theoretical uncertainties are obtained by summing scale, PDF, and $\alpha_{s}$ uncertainties in quadrature. Then, total uncertainties are symmetrized by taking the bigger values from estimated up and down uncertainties in a conservative approach.  

\begin{table}
\caption{The estimated sizes of the scale, PDF, and $\alpha_{s}$ uncertainties for the 8 TeV inclusive cross section predictions of the W$^{+}$ and W$^{-}$ bosons at NNLO accuracy. Estimated uncertainties are given as percentage (\%) of the central values of the predictions from NNPDF3.1, CT14, and MMHT2014 PDF sets.}
\label{tab:1}    
\begin{tabular}{lccc}
\hline\noalign{\smallskip}
Uncertainty  & NNPDF3.1 & CT14 & MMHT2014  \\
\noalign{\smallskip}\hline\noalign{\smallskip}
\multicolumn{4}{c}{Values for $W^{+}\rightarrow e^{+}\nu$ process}   \\
Scale (\%)  & 0.74     & 0.76        & 0.78    \\
PDF (\%)    & 1.96     & 2.40        & 1.64    \\
$\alpha_{s}$ (\%)      & 1.06         & 1.04        & 1.10     \\
Total (\%)   & 2.35     & 2.72         & 2.12     \\
\noalign{\smallskip}\hline\noalign{\smallskip}
\multicolumn{4}{c}{Values for $W^{-}\rightarrow e^{-}\nu$ process}   \\
Scale (\%) & 0.72     & 0.64        & 0.80    \\
PDF (\%)    & 2.22     & 2.90        & 1.50    \\
$\alpha_{s}$ (\%)    & 1.16         & 1.00        & 1.14     \\
Total (\%)   & 2.61     & 3.13         & 2.05     \\
\noalign{\smallskip}\hline
\end{tabular}
\end{table}

\section{Phenomenological results at 8 TeV }
\label{8tev}
The NNLO calculations and their comparisons with the reference LHCb results at 8 TeV~\cite{Aaij:2016qqz} are reported in this section. The calculations are performed in the fiducial phase space as discussed in Sec.~\ref{fid}. W$^{+}$/W$^{-}$ differential cross section ratio $R_{W^{\pm}}$, where

\begin{equation}
\label{eqn:3}   
R_{W^{\pm}}=\frac{d\sigma(W^{+}\rightarrow e^{+}\nu)/d\eta_{e}}{d\sigma(W^{-}\rightarrow e^{-}\nu)/d\eta_{e}},
\end{equation}
and electron charge asymmetry $A_{e}$ (Eq.~\ref{eqn:2}) variables as a function of $\eta_{e}$ are used in justification of the calculations. The $R_{W^{\pm}}$ and $A_{e}$ variables are calculated in the $\eta_{e}$ bins of (2.0, 2.25), (2.25, 2.5), (2.5, 2.75), (2.75, 3.0), (3.0, 3.25), (3.25, 3.5), (3.5, 3.75), (3.75, 4.25) in line with the 8 TeV LHCb paper. The NNPDF3.1, CT14, and MMHT2014 NNLO PDF sets are used in the calculations. Total theoretical uncertainties that are estimated as discussed in Sec.~\ref{theo} are propagated to the $R_{W^{\pm}}$ and $A_{e}$ calculations. The LHCb data results that were obtained at Born level, do not incorporate the effect of quantum electrodynamics final-state radiation, are used to enable direct comparisons to the calculations. The data central results in the comparisons are used together with their total experimental uncertainties that are obtained by adding statistical, systematic, and the LHC beam energy uncertainties in quadrature. 

\begin{figure}
\includegraphics[width=9.25cm]{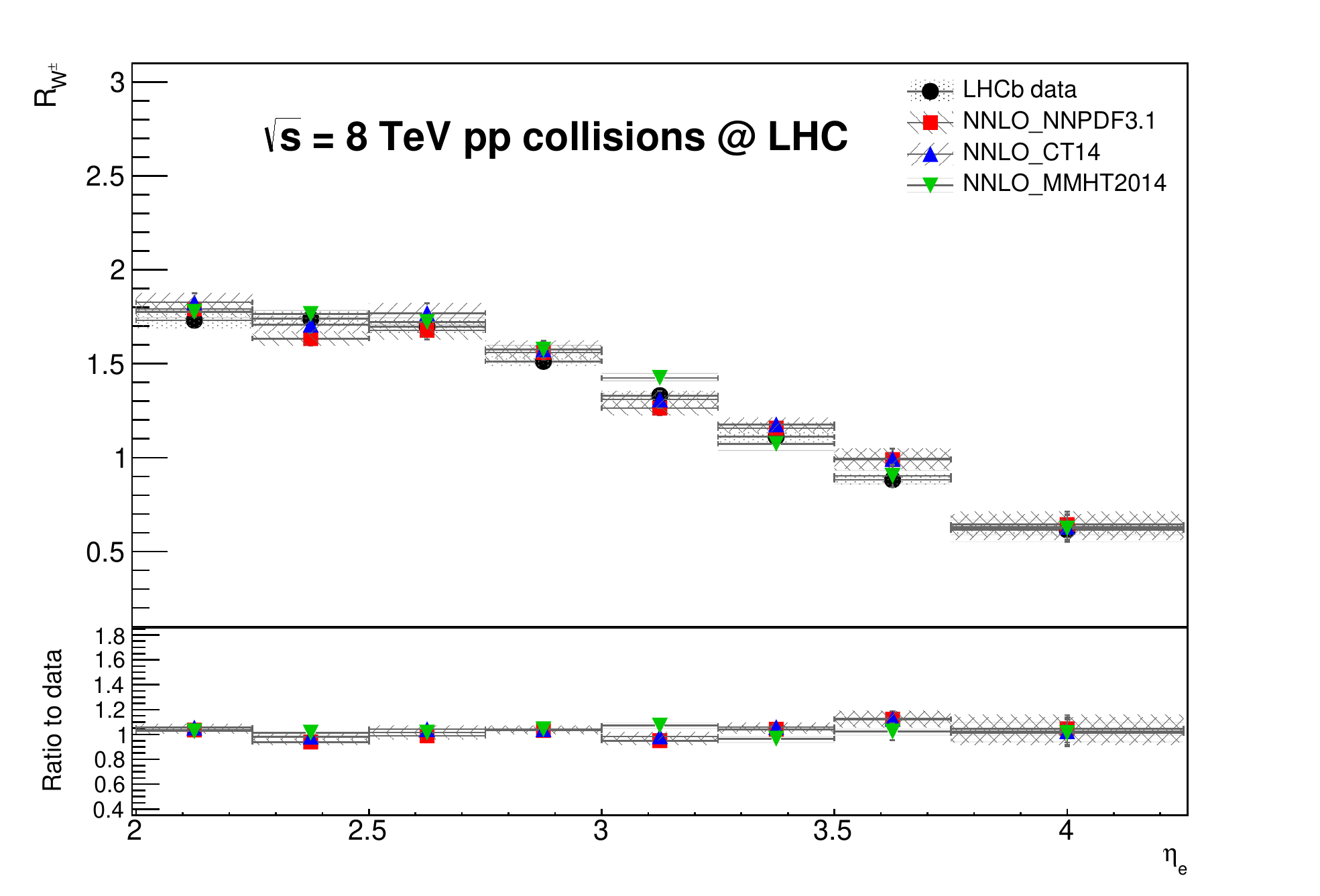}
\caption{The differential cross section ratio $R_{W^{\pm}}$ distributions as a function of the $\eta_{e}$ which are predicted at NNLO accuracy by using NNPDF3.1, CT14, and MMHT2014 PDF sets and their comparisons with the LHCb data at 8 TeV. The predictions include total theoretical uncertainties obtained by adding scale, PDF, and $\alpha_s$ uncertainties in quadrature. The data includes total experimental uncertainty obtained by adding statistical, systematic, and the LHC beam energy uncertainties in quadrature. In the lower inset, the ratios of the predictions to the data for the $R_{W^{\pm}}$ variable are provided.}
\label{fig:1}      
\end{figure}

The predicted $R_{W^{\pm}}$ distributions based on different PDF sets are compared with the data in Fig.~\ref{fig:1}. The $R_{W^{\pm}}$ predictions at NNLO accuracy using different PDF sets are generally in good agreement with each other and with the data within uncertainties through the entire $\eta_{e}$ bins. The prediction using MMHT2014 PDF set provides the best agreement with the data among other predictions except for the 3.0--3.25 bin where it overestimates the data by up to 5\%. The predictions using NNPDF3.1 and CT14 overestimate the data in 3.5--3.75 bin by up to 10\%, while the prediction using MMHT2014 is in quite good agreement with the data in this bin. The predictions using different PDF sets are able to describe the data well in the very forward $\eta_{e}$ bin 3.75--4.25. It is worth to notice that the $\emph{FEWZ}$ predictions provided in the LHCb paper tend to underestimate the $R_{W^{\pm}}$ data distribution in the far forward bins 3.5--3.75 and 3.75--4.25. The discrepancy between the \emph{MATRIX} and $\emph{FEWZ}$ predictions is checked and attributed mainly to the residual dependence of the  $r_{cut}$ = 0.15\% value of the \emph{$q_{T}$}-subtraction method in the far forward region. The \emph{MATRIX} framework differs by the \emph{$q_{T}$}-subtraction method from the $\emph{FEWZ}$ which is based on sector decomposition~\cite{Binoth:2000ps}. This residual dependence is due to power-suppressed terms that remain after the subtraction of infrared singular contribution at finite $r_{cut}$ value. The impact of the choice of $r_{cut}$ = 0.15\% is checked with respect to a lower cut-off value 0.05\%, which amounts to $\sim$2--7\% difference. This difference shows that the power corrections due to finite $r_{cut}$ = 0.15\% value are large in this region which can only vanish in the limit  $r_{cut} \rightarrow$ 0. A systematic uncertainty up to 4\% is propagated to the $R_{W^{\pm}}$ predictions in the last two bins from the extrapolation $r_{cut} \rightarrow$ 0 to account for this dependency to the $r_{cut}$ value by means of the large power corrections. In addition, the impact of the difference in the input parameter choices of the \emph{MATRIX} and the $\emph{FEWZ}$ default setup~\cite{Li:2012wna} is checked to be 1--2\%. Together with these checks, the NNLO calculations in terms of the predicted $R_{W^{\pm}}$ distributions are therefore justified quite well by using the LHCb results.                              

\begin{figure}
\includegraphics[width=9.25cm]{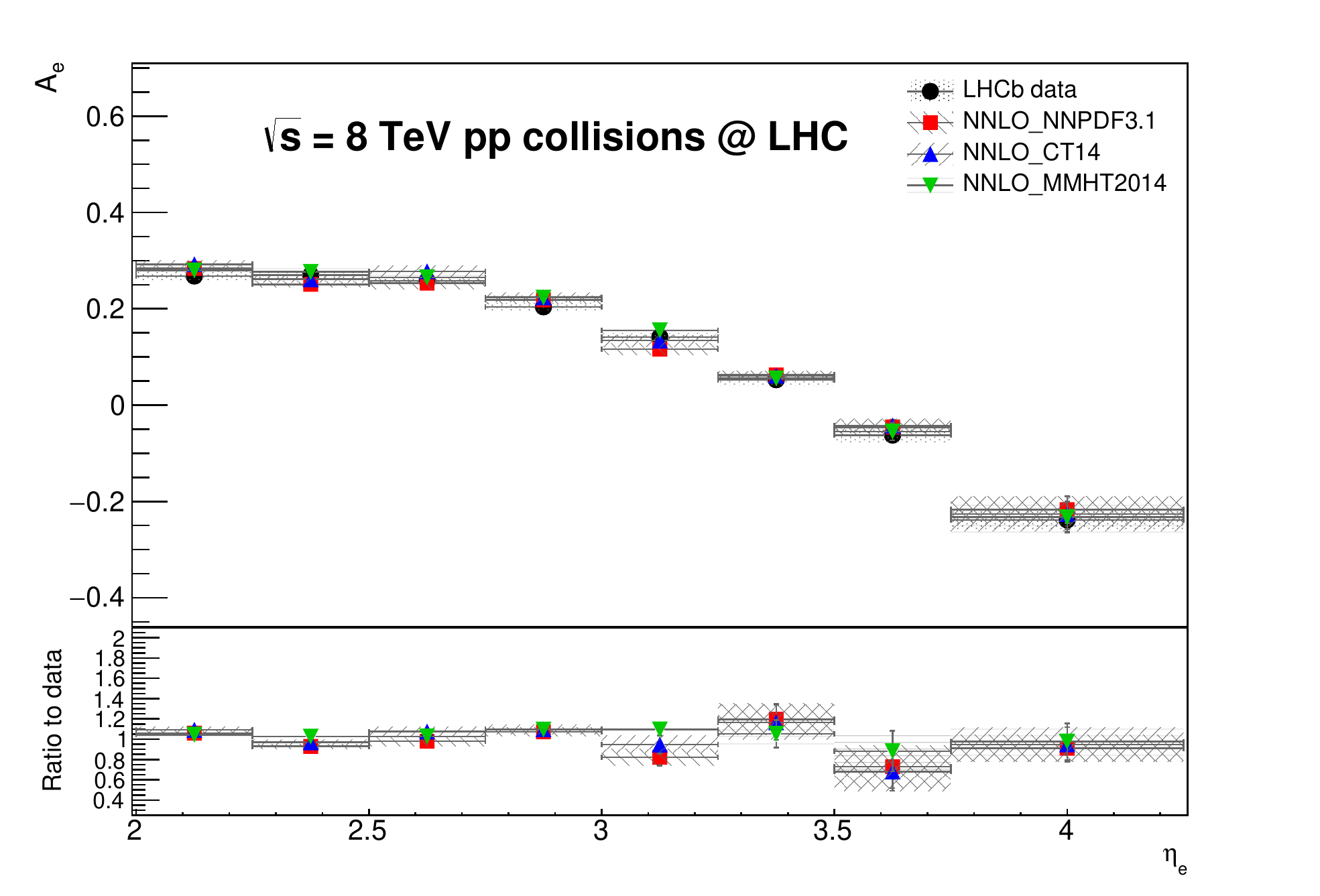}
\caption{The electron charge asymmetry $A_{e}$ distributions as a function of the $\eta_{e}$ which are predicted at NNLO accuracy by using NNPDF3.1, CT14, and MMHT2014 PDF sets and their comparisons with the LHCb data at 8 TeV. The predictions include total theoretical uncertainties while the data include total experimental uncertainty. In the lower inset, the ratios of the predictions to the data for the $A_{e}$ variable are provided.}
\label{fig:2}      
\end{figure}

The predictions from different PDF sets and their comparisons with the data at NNLO accuracy for the $A_{e}$ variable are shown in Fig.~\ref{fig:2} and also tabulated in Table~\ref{tab:2}. The predictions using different PDF sets generally reproduce the data well within uncertainties throughout the entire $\eta_{e}$ ranges. The predictions are also in good agreement with each other and with the data except for the bins 3.0--3.25 and 3.5--3.75, where the predictions using NNPDF3.1 and CT14 exhibit lower values as compared to the data and the prediction using MMHT2014. The discrepancy between the data and the predictions using NNPDF3.1 and CT14 are up to $\sim$15\% in the bins 3.0--3.25 and 3.5--3.75. The best description of the data is achieved with the prediction using MMHT2014 over the other PDF sets for the entire $\eta_{e}$ ranges. The predictions using different PDF sets are able to describe the data in the far forward bins 3.5--3.75 and 3.75--4.25 in comparison to the $\emph{FEWZ}$ predictions provided in the LHCb paper, where they slightly tend to underestimate the data. More generally the trend of increasing discrepancy between the data and predictions towards the very forward $\eta_{e}$ region as presented in the LHCb paper is not observed in Fig.~\ref{fig:2} for the central values of the predictions. This discrepancy between the \emph{MATRIX} and $\emph{FEWZ}$ predictions for the description of the data in the far forward bins is also mainly due to the large power corrections that appear at finite cut-off $r_{cut}$ = 0.15\% value in the MATRIX calculation. The difference of the choice of $r_{cut}$ = 0.15\% relative to a lower cut-off value $r_{cut}$ = 0.05\% has been checked to be $\sim$14--19\%. Therefore, a systematic uncertainty up to 5\% from the extrapolation in the limit $r_{cut} \rightarrow$ 0 is also propagated to the predictions in the last two bins of the $A_{e}$ variable to account for the effect from the large power corrections at the finite $r_{cut}$ value in the subtraction method. To this end, the NNLO calculations using different PDF sets are successfully justified with the LHCb results for also the $A_{e}$ variable at 8 TeV.    

\begin{table}
\caption{The 8 TeV electron charge asymmetry (in percent) $A_{e}$(\%) predictions by using NNPDF3.1, CT14, and MMHT2014 PDF sets at NNLO accuracy and their comparisons with the LHCb data in bins of the $\eta_{e}$. The predictions include total theoretical uncertainties while the data include total experimental uncertainty.}
\label{tab:2}    
\begin{scriptsize}   
\begin{tabular}{ccccc}
\hline\noalign{\smallskip}
$\eta_{e}$ & Data & NNPDF & CT & MMHT  \\
\noalign{\smallskip}\hline\noalign{\smallskip}
2.00--2.25  & 26.78$\pm$0.8     & 28.37$\pm$0.7        & 29.24$\pm$0.8     & 27.95$\pm$0.6  \\
2.25--2.50  & 26.98$\pm$0.7     & 25.06$\pm$0.6        & 26.14$\pm$0.7     & 27.66$\pm$0.7  \\
2.50--2.75  & 25.84$\pm$0.7     & 25.30$\pm$1.3        & 27.78$\pm$1.2     & 26.53$\pm$1.0  \\
2.75--3.00  & 20.39$\pm$0.8     & 21.85$\pm$0.8        & 22.37$\pm$0.9     & 22.31$\pm$0.9  \\
3.00--3.25  & 14.15$\pm$0.8     & 11.62$\pm$1.1        & 13.40$\pm$1.2     & 15.50$\pm$0.8  \\
3.25--3.50  & 5.25$\pm$1.2       & 6.28$\pm$0.8          & 6.12$\pm$0.9       & 5.53$\pm$1.0  \\
3.50--3.75  & -6.25$\pm$1.4      & -4.55$\pm$1.5         & -4.26$\pm$1.5     & -5.51$\pm$1.8  \\
3.75--4.25  & -23.85$\pm$1.9    & -21.70$\pm$2.8       & -22.61$\pm$3.6   & -23.30$\pm$3.2  \\
\noalign{\smallskip}\hline
\end{tabular}
\end{scriptsize}
\end{table}

\section{Predictions at 13 TeV and 14 TeV}
\label{13tev}
The predictions from the NNLO calculations are validated by using the LHCb data at 8 TeV in Sec.~\ref{8tev}. The 8 TeV predictions using all the PDF sets, in particular the MMHT2014 PDF set, described the data well in a consistent manner throughout the entire $\eta_{e}$ region and already motivated extensions of the calculations to 13 TeV and 14 TeV, which are the current and the near-future planned center-of-mass energies of the LHC, respectively. The 13 TeV and 14 TeV NNLO calculations are performed by employing the same fiducial phase space requirements as discussed in Sec.~\ref{fid}. Total theoretical uncertainties are estimated using the procedure as detailed in Sec.~\ref{theo}. Additionally, a systematic uncertainty is included for the predictions in the last two forward bins to account for the dependency to the finite cut-off $r_{cut}$ = 0.15\% value, where the large power corrections can have an impact as already discussed in Sec.~\ref{8tev}. The PDF sets NNPDF3.1, CT14, and MMHT2014 are used in the calculations.      

\begin{figure}
\includegraphics[width=9.2cm]{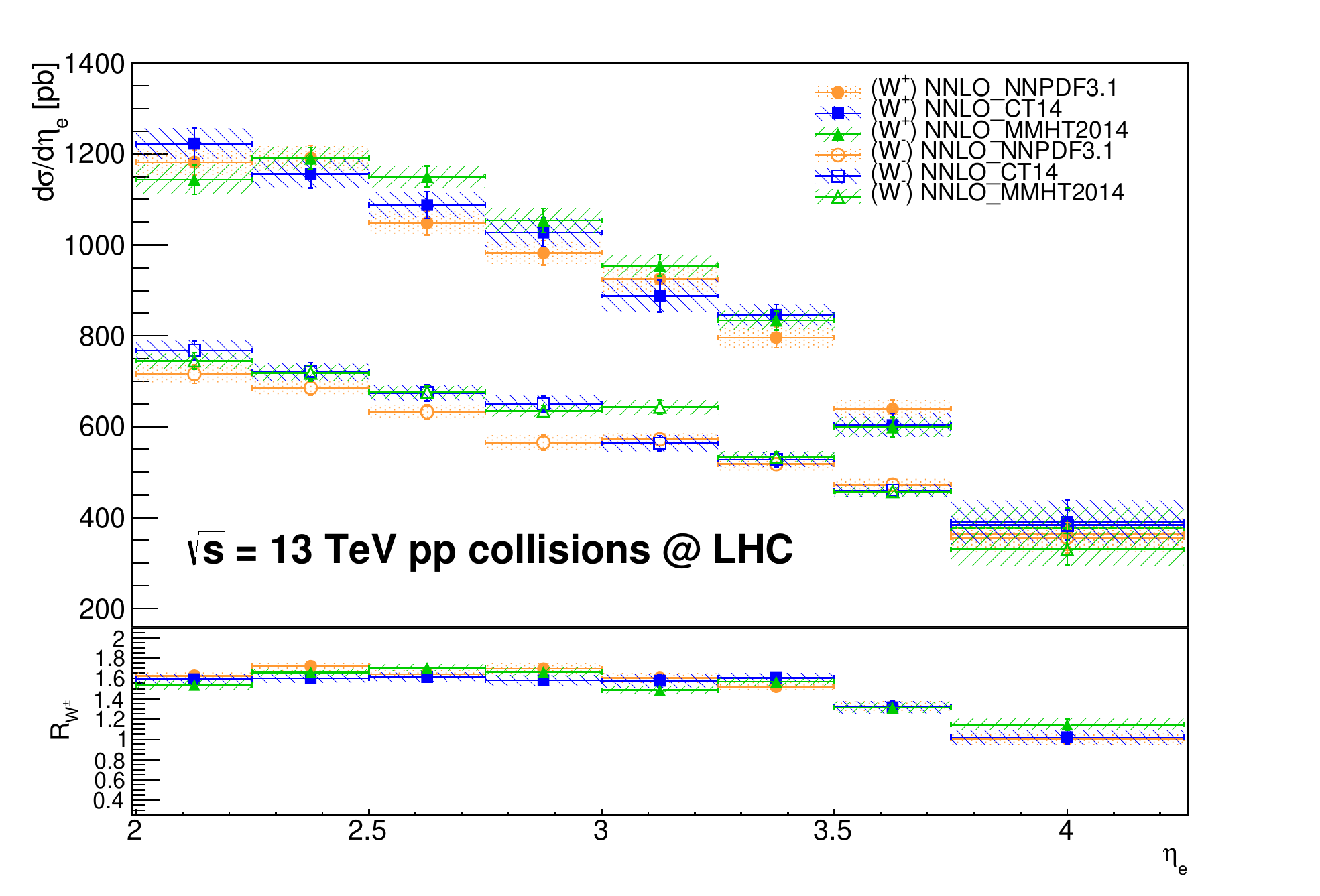}
\includegraphics[width=9.2cm]{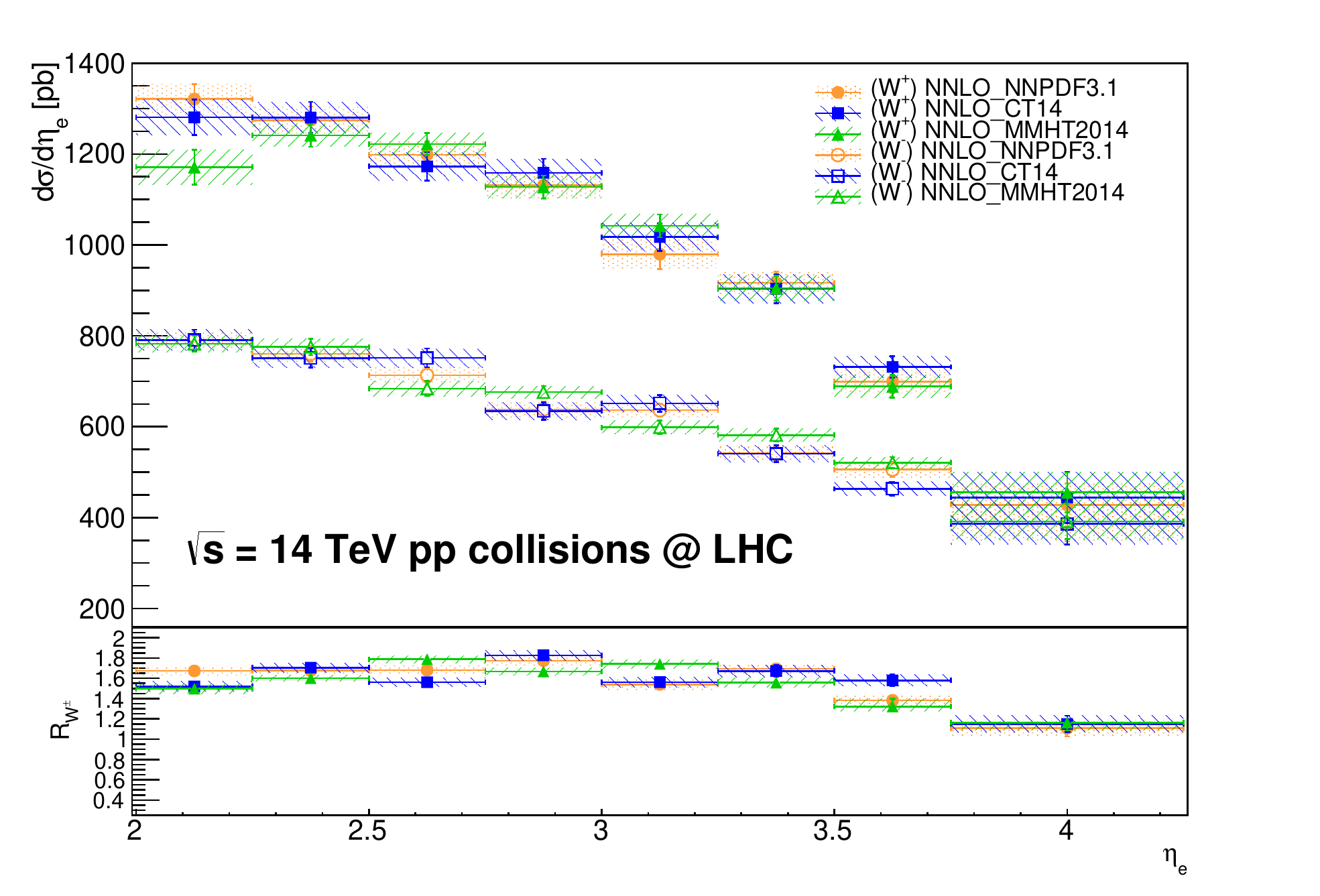}

\caption{The 13 TeV (top) and 14 TeV (bottom) differential cross section $d\sigma$ distributions for the W$^{+}$ and W$^{-}$ boson processes as a function of the $\eta_{e}$ which are predicted at NNLO accuracy by using NNPDF3.1, CT14, and MMHT2014 PDF sets. The predictions include total theoretical uncertainties. In the lower inset, the predictions for the differential cross section ratios $R_{W^{\pm}}$ are provided.}
\label{fig:3}      
\end{figure}

The 13 TeV and 14 TeV calculations are first performed for the predictions of the differential cross sections of the W$^{\pm}$ bosons $d\sigma(W^{\pm})$ and their ratios $R_{W^{\pm}}$ in bins of the $\eta_{e}$. The differential distributions are predicted and compared among different PDF models in Fig.~\ref{fig:3}. The 13 TeV differential distributions are predicted consistently within uncertainties for the W$^{\pm}$ boson processes only with a few exceptions where the predictions using different PDF sets exhibited slight deviations from each other such as in bins 2.75--3.0 and 3.0--3.25 for the W$^{-}$ boson process. The prediction using NNPDF3.1 slightly underestimates the W$^{-}$ boson cross section for the lower $\eta_{e}$ bins with respect to the predictions using other PDF sets. However, the predicted $R_{W^{\pm}}$ distributions are in very good agreement among different PDF models. The $R_{W^{\pm}}$ distributions tend to decrease towards forward $\eta_{e}$ bins as expected. The 14 TeV differential cross sections are also predicted consistently within uncertainties for the W$^{\pm}$ boson processes regardless of the PDF model. The predictions only exhibit small discrepancies that are more pronounced in bins 3.25--3.5 and 3.5--3.75. The predicted 14 TeV $R_{W^{\pm}}$ distributions are also in good agreement among PDF sets, however become more sensitive to distinguish among different PDF models as compared to 13 TeV distributions. This is clearly seen in the intermediate to higher bins of $\eta_{e}$ that the predictions tend to deviate from each other by up to $\sim$12\%. Therefore, the NNLO differential cross section distributions for the W$^{\pm}$ boson processes can be further used to calculate $A_{e}$ distributions at 13 TeV and 14 TeV.                  

\begin{figure}
\includegraphics[width=9.25cm]{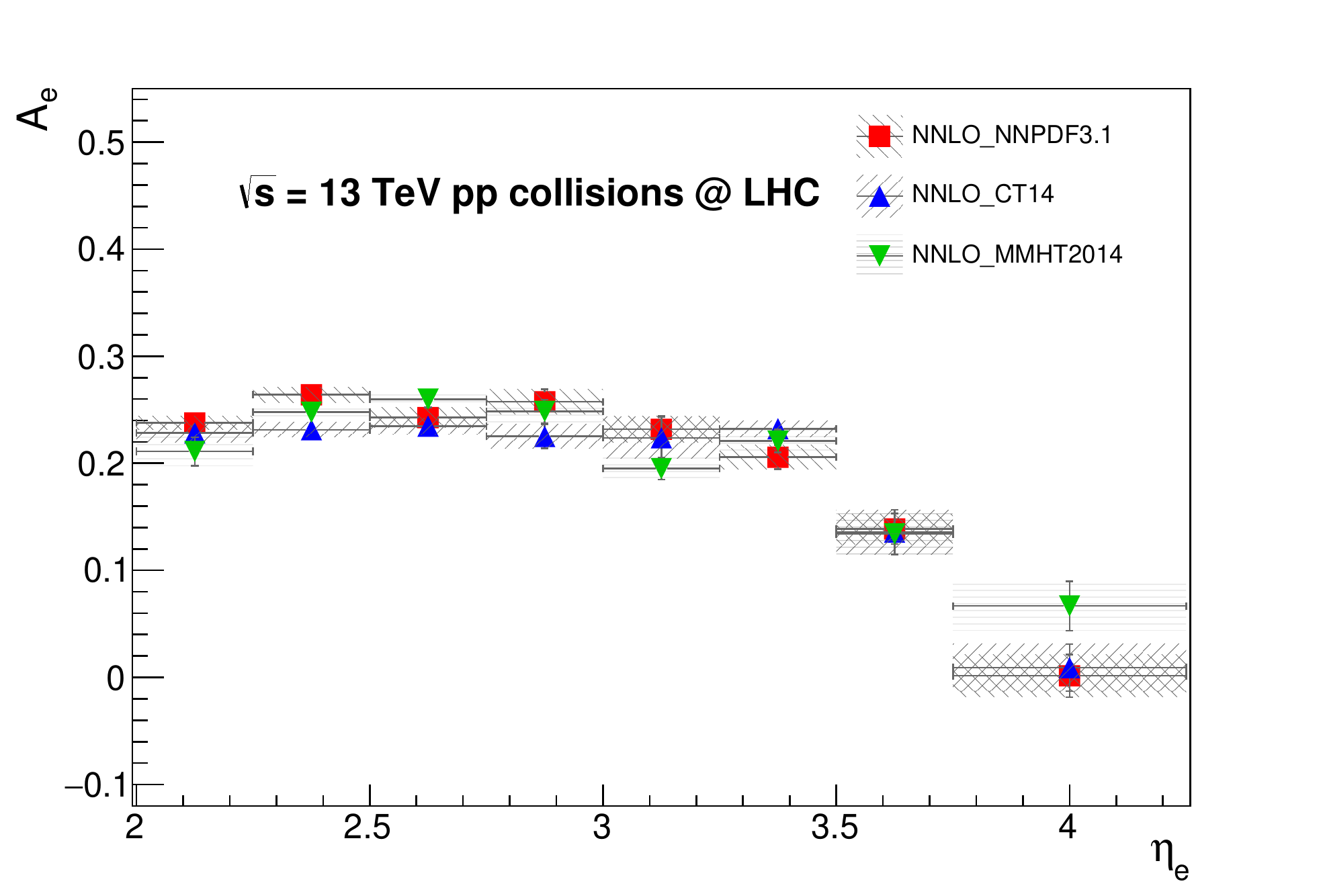}
\includegraphics[width=9.25cm]{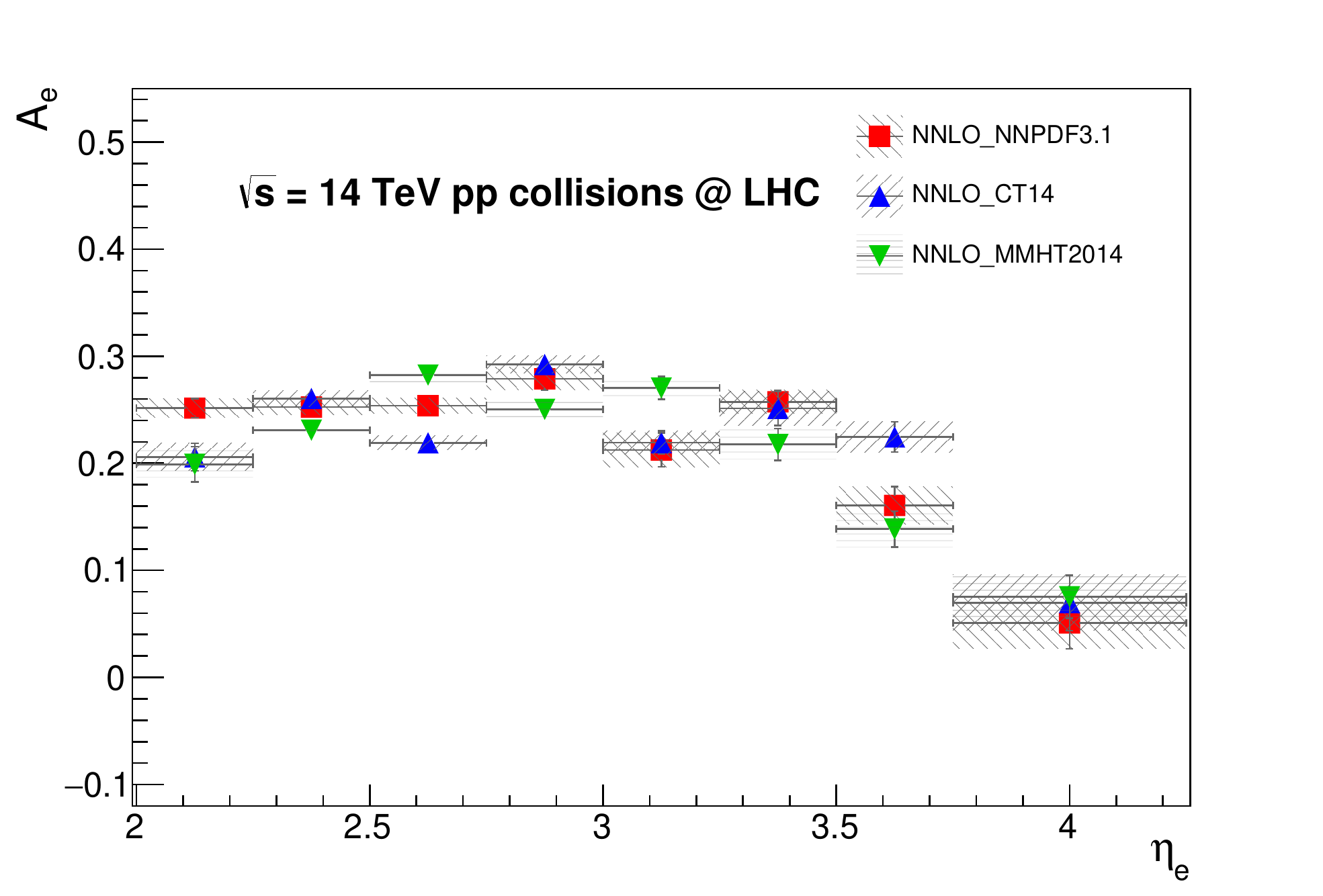}

\caption{The 13 TeV (top) and 14 TeV (bottom) electron charge asymmetry $A_{e}$ distributions as a function of the $\eta_{e}$ which are predicted at NNLO accuracy by using NNPDF3.1, CT14, and MMHT2014 PDF sets. The predictions include total theoretical uncertainties.}
\label{fig:4}      
\end{figure}

Next the 13 TeV and 14 TeV electron charge asymmetry $A_{e}$ calculations are performed as a function of the $\eta_{e}$ based on the differential cross section predictions of the W$^{\pm}$ boson processes. The differential $A_{e}$ predictions are compared among the PDF sets as shown in Fig.~\ref{fig:4} and tabulated in Tables~\ref{tab:3} and~\ref{tab:4}. Overall the 13 TeV predictions for the $A_{e}$ variable from different PDF sets are in good agreement with each other within uncertainties almost for the entire $\eta_{e}$ ranges. The $A_{e}$ predictions in the most forward $\eta_{e}$ bin 3.75--4.25 distinguish MMHT2014 PDF set from other PDF sets, where the prediction using MMHT2014 provides much higher $A_{e}$ result. The 14 TeV predictions for the $A_{e}$ variable are mostly in good agreement among the PDF sets within uncertainties, however more sensitive to distinguish among them as compared to the 13 TeV predictions. The predictions using NNPDF3.1 and CT14 agree with each other more than the prediction using MMHT2014. The prediction using CT14 exhibits some deviations that are more pronounced in intermediate region. No single PDF set is able to provide consistent agreement with the other throughout the entire $\eta_{e}$ ranges. The sensitivity to the PDF choice regarding the $A_{e}$ distributions is clearly increased in going from 13 TeV results to 14 TeV results.                    

\begin{table}
\caption{The 13 TeV electron charge asymmetry (in percent) $A_{e}$(\%) predictions by using NNPDF3.1, CT14, and MMHT2014 PDF sets at NNLO accuracy in bins of the $\eta_{e}$. The predictions include total theoretical uncertainties.}
\label{tab:3}    
\begin{tabular}{cccc}
\hline\noalign{\smallskip}
$\eta_{e}$ & NNPDF3.1 & CT14 & MMHT2014  \\
\noalign{\smallskip}\hline\noalign{\smallskip}
2.00--2.25  & 23.78$\pm$0.6     & 22.84$\pm$0.9        & 21.11$\pm$1.3     \\
2.25--2.50  & 26.40$\pm$0.7     & 23.13$\pm$0.7        & 24.77$\pm$0.6     \\
2.50--2.75  & 24.27$\pm$1.0     & 23.47$\pm$0.7        & 26.00$\pm$1.6     \\
2.75--3.00  & 25.75$\pm$1.2     & 22.51$\pm$1.1        & 24.85$\pm$1.1     \\
3.00--3.25  & 23.18$\pm$1.2     & 22.38$\pm$1.9        & 19.50$\pm$1.0     \\
3.25--3.50  & 20.57$\pm$1.1     & 23.23$\pm$0.8        & 22.07$\pm$1.1     \\
3.50--3.75  & 13.86$\pm$1.7     & 13.57$\pm$1.9        & 13.41$\pm$1.9     \\
3.75--4.25  & 0.15$\pm$2.0       & 0.91$\pm$2.2          & 6.67$\pm$2.3       \\
\noalign{\smallskip}\hline
\end{tabular}
\end{table}

\begin{table}
\caption{The 14 TeV electron charge asymmetry (in percent) $A_{e}$(\%) predictions by using NNPDF3.1, CT14, and MMHT2014 PDF sets at NNLO accuracy in bins of the $\eta_{e}$. The predictions include total theoretical uncertainties.}
\label{tab:4}    
\begin{tabular}{cccc}
\hline\noalign{\smallskip}
$\eta_{e}$ & NNPDF3.1 & CT14 & MMHT2014  \\
\noalign{\smallskip}\hline\noalign{\smallskip}
2.00--2.25  & 25.15$\pm$0.9     & 20.60$\pm$1.3        & 19.90$\pm$1.6       \\
2.25--2.50  & 25.26$\pm$0.7     & 26.05$\pm$0.8        & 23.09$\pm$0.5       \\
2.50--2.75  & 25.38$\pm$0.7     & 21.91$\pm$0.7        & 28.24$\pm$0.6     \\
2.75--3.00  & 27.90$\pm$1.0     & 29.22$\pm$0.8        & 25.04$\pm$0.8     \\
3.00--3.25  & 21.24$\pm$1.5     & 21.93$\pm$1.1        & 27.04$\pm$1.1      \\
3.25--3.50  & 25.74$\pm$1.1     & 25.13$\pm$1.6        & 21.76$\pm$1.5         \\
3.50--3.75  & 16.06$\pm$1.8     & 22.46$\pm$1.4        & 13.87$\pm$1.7      \\
3.75--4.25  & 5.08$\pm$2.4       & 6.97$\pm$2.6          & 7.53$\pm$2.0    \\
\noalign{\smallskip}\hline
\end{tabular}
\end{table}

\begin{figure}
\includegraphics[width=9.25cm]{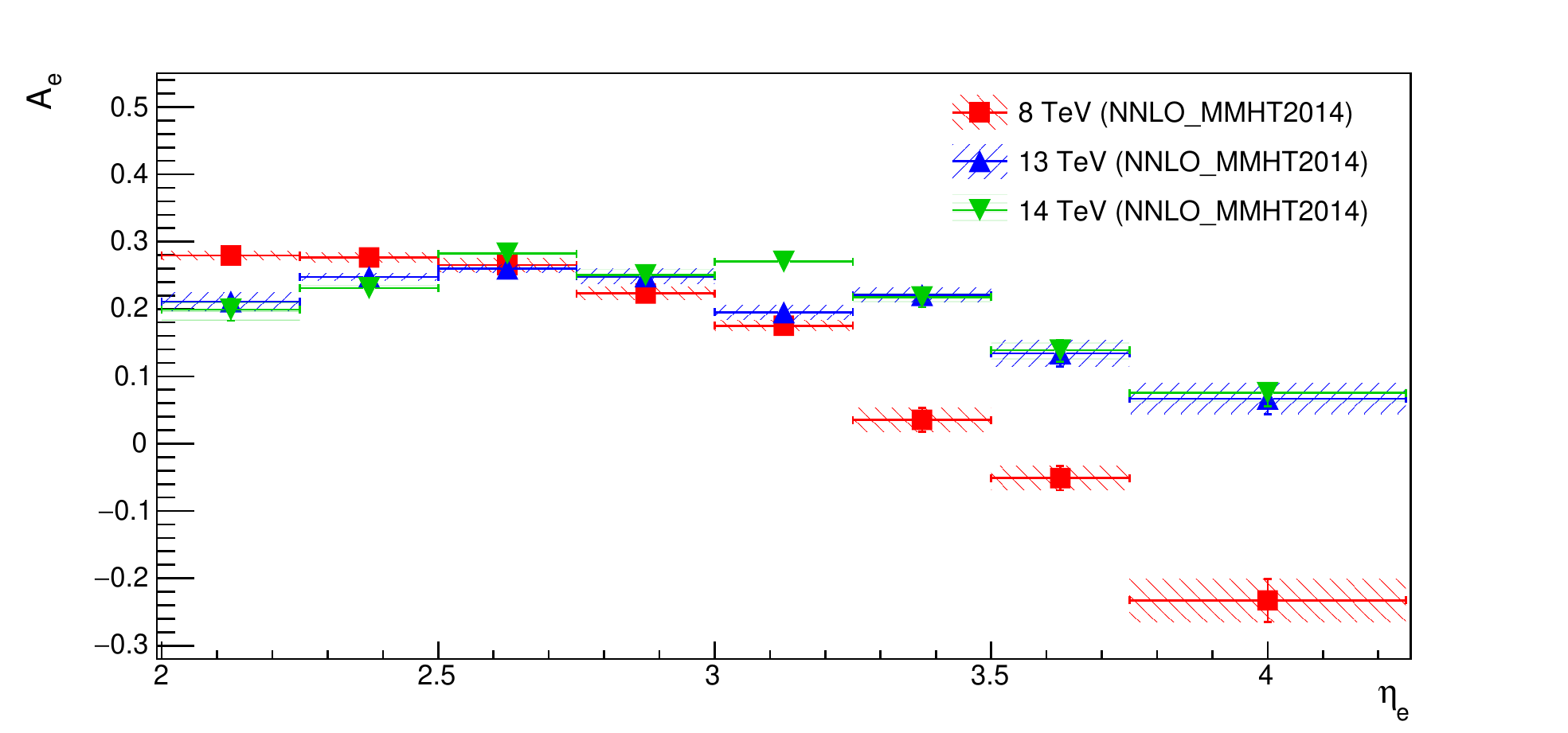}
\caption{The comparison of the electron charge asymmetry $A_{e}$ distributions at 8 TeV, 13 TeV, and 14 TeV as a function of the $\eta_{e}$ which are predicted by using MMHT2014 PDF set. The prediction includes total theoretical uncertainties.}
\label{fig:5}      
\end{figure}

The $A_{e}$ distributions are also compared at different center-of-mass energies 8 TeV, 13 TeV, and 14 TeV for the prediction using MMHT2014 as a baseline PDF set in Fig.~\ref{fig:5}. It has been discussed in Sec.~\ref{8tev} that the prediction using MMHT2014 provides the best description of 8 TeV LHCb data and also provides $A_{e}$ distributions reasonably well at 13 TeV and 14 TeV. The 13 TeV and 14 TeV predictions provided lower distributions in the lower $\eta_{e}$ bins 2.0--2.5 and higher distributions in the intermediate to higher bins 2.75--4.25 relative to the 8 TeV prediction. The distributions at different energies almost overlap within uncertainties in the bins 2.5--3.0. Right after the bin 2.75--3.0, the 13 TeV and 14 TeV predictions separate from the 8 TeV prediction by exhibiting higher distributions towards the very forward $\eta_{e}$ region. In the most forward bins 3.25--4.25, 13 TeV and 14 TeV predictions overlap within uncertainties, however 14 TeV distribution is predicted slightly higher than 13 TeV distribution. The charge asymmetry increases by going from 8 TeV to higher energies in the very forward $\eta_{e}$ bins 3.0--4.25. Another outstanding observation is that $\eta_{e}$ bin where $A_{e}$ distribution reaches a peak shifts towards higher bins in going from 8 TeV to higher energies.    
  
\section{Summary and conclusion}
\label{conclusion}
In this paper a detailed study of the theoretical predictions of the electron charge asymmetry for the inclusive $W^{\pm}+X \rightarrow e^{\pm} \nu +X$ production in the forward region of the pp collisions is presented. The charge asymmetry calculations are performed with the inclusion of QCD NNLO corrections in the perturbation expansion at 8 TeV, 13 TeV, and 14 TeV LHC energies. The calculations are considered for the fiducial acceptance of the decay electron with transverse momentum $p_{T}>$ 20 GeV in the forward electron pseudorapidity region $2.0 \leq \eta_{e} \leq 4.25$. The charge asymmetry in the forward $\eta_{e}$ region is sensitive to the ratio of $u$ and $d$ quark distributions in the proton as a function of Bjorken-$x$ values of the partons, particularly providing valuable inputs for the accurate determination of PDFs at small and large $x$ values between $10^{-4}\leq x \leq 10^{-1}$.   

The NNLO predictions that are based on the PDF sets NNPDF3.1, CT14, and MMHT2014 at 8 TeV are first presented and compared with the reference LHCb measurement~\cite{Aaij:2016qqz}. The 8 TeV predictions are obtained for the W$^{+}$/W$^{-}$ cross section ratio and the charge asymmetry differential in bins of the $\eta_{e}$. The predicted results from different PDF sets are found to be in good agreement with each other and with the LHCb data for most of the $\eta_{e}$ bins. The results showed that the best description of the 8 TeV data is achieved by the prediction using MMHT2014 for both the cross section ratio and charge asymmetry. Furthermore, the predictions from different PDF sets are able to describe the data quite well in the far forward $\eta_{e}$ bins 3.5--3.75 and 3.75--4.25 in comparison to the predictions provided with the LHCb measurement, where they tend to underestimate the data within theoretical and experimental uncertainties. The NNLO predictions using different PDF sets are therefore successfully justified with the LHCb data at 8 TeV.   

The 13 TeV and 14 TeV NNLO results that are presented in this paper represent the first NNLO predictions of the charge asymmetry for the W$^{\pm}$ boson processes in the forward region $2.0 \leq \eta_{e} \leq 4.25$. The 13 TeV and 14 TeV predictions are reported for the differential cross sections of the W$^{\pm}$ bosons and the charge asymmetry as a function of the $\eta_{e}$. The predicted results from NNPDF3.1, CT14, and MMHT2014 are compared to test the sensitivity to discriminate among different PDF models. The W$^{+}$ and W$^{-}$ boson differential cross sections are predicted consistently among the PDF sets apart from some $\eta_{e}$ bins where the predictions from different PDF models exhibited slight discrepancies with respect to each other. The 14 TeV predictions for the differential cross sections and their ratios are found to be more sensitive to distinguish among the PDF models in comparison to the 13 TeV predictions. The NNLO charge asymmetry results are reported to be in good agreement within total theoretical uncertainties among the predictions using different PDF sets for almost the entire $\eta_{e}$ ranges at 13 TeV and 14 TeV. In addition, the sensitivity of the charge asymmetry distributions to the choice of PDF set is shown to be increased with the increasing collision energy in going from 13 TeV to 14 TeV. Moreover, The predicted charge asymmetry results are compared at 8 TeV, 13 TeV, and 14 TeV collision energies by using MMHT2014 as a baseline PDF set. The 13 TeV and 14 TeV distributions are shown to separate from 8 TeV distribution in the $\eta_{e}$ bin 3.0--3.25 towards the very forward $\eta_{e}$ region, where the charge asymmetry is increased in going from 8 TeV to 13 TeV and 14 TeV distributions. In the forward bins $\eta_{e}$ 3.25--4.25, the 13 TeV and 14 TeV distributions are observed to overlap within uncertainties, however 14 TeV charge asymmetry distribution is predicted to be slightly higher than 13 TeV distribution. The results clearly show that the very forward $\eta_{e}$ bins become more of interest to probe precisely the $u$ and $d$ quark distributions in the proton by shifting towards 13 TeV and 14 TeV collision energies.                      


\section*{Conflict of interest}
The author declares that he has no conflict of interest.



\begin{thebibliography}{}
\bibitem{Abe:1998rv}
F. Abe et al., Measurement of the lepton charge asymmetry in W boson decays produced in p$\rm{\bar{p}}$ collisions, Phys. Rev. Lett., 81, 5754-5759 (1998)
\bibitem{Abazov:2007pm}
V. M. Abazov et al., Measurement of the muon charge asymmetry from W boson decays, Phys. Rev. D, 77, 011106 (2008)
\bibitem{Abazov:2008qv}
V. M. Abazov et al., Measurement of the electron charge asymmetry in $\rm{p\rm{\bar{p}} \rightarrow W+X \rightarrow e\nu+X}$ events at $\sqrt{s}$ = 1.96 TeV, Phys. Rev. Lett., 101, 211801 (2008)
\bibitem{Aaltonen:2009ta}
T. Aaltonen et al., Direct measurement of the W production charge asymmetry in p$\rm{\bar{p}}$ collisions at $\sqrt{s}$ = 1.96 TeV, Phys. Rev. Lett., 102, 181801 (2009)
\bibitem{Abazov:2013rja}
V. M. Abazov et al., Measurement of the muon charge asymmetry in $\rm{p\rm{\bar{p}} \rightarrow W+X \rightarrow \mu\nu+X}$ events at $\sqrt{s}$ = 1.96 TeV, Phys. Rev. D, 88, 091102 (2013)

\bibitem{Aad:2011dm}
G. Aad et al., Measurement of the inclusive $W^{\pm}$ and $Z/\gamma$ cross sections in the electron and muon decay channels in pp collisions at $\sqrt{s}$ = 7 TeV with the ATLAS detector, Phys. Rev. D, 85, 072004 (2012)
\bibitem{Chatrchyan:2012xt}
S. Chatrchyan et al., Measurement of the electron charge asymmetry in Inclusive W production in pp collisions at $\sqrt{s}$ = 7 TeV, Phys. Rev. Lett., 109, 111806 (2012)
\bibitem{Chatrchyan:2013mza}
S. Chatrchyan et al., Measurement of the muon charge asymmetry in Inclusive $pp\rightarrow W+X$ production at $\sqrt{s}$ = 7 TeV and an improved determination of light parton distribution functions, Phys. Rev. D, 90, 032004 (2014)
\bibitem{Khachatryan:2016pev}
V. Khachatryan et al., Measurement of the differential cross section and charge asymmetry for inclusive $pp\rightarrow W^{\pm}+X$ production at $\sqrt{s}$ = 8 TeV, Eur. Phys. J. C, 76, 469 (2016)
\bibitem{Aaboud:2016btc}
M. Aaboud et al., Precision measurement and interpretation of inclusive $W^{+}$, $W^{-}$, and $Z/\gamma^{\star}$ production cross sections with the ATLAS detector, Eur. Phys. J. C, 77, 367 (2017)
\bibitem{Aad:2019bdc}
G. Aad et al., Measurement of $W^{\pm}$-boson and $Z$-boson production cross-sections in pp collisions at $\sqrt{s}$ = 2.76 TeV with the ATLAS detector, Eur. Phys. J. C, 79, 901 (2019)
\bibitem{Aaboud:2018nic}
M. Aaboud et al., Measurements of $W$ and $Z$ boson production in pp collisions at $\sqrt{s}$ = 5.02 TeV with the ATLAS detector, Eur. Phys. J. C, 79, 128 (2019)
\bibitem{Aad:2019rou}
G. Aad et al., Measurement of the cross-section and charge asymmetry of $W$ bosons produced in proton-proton collisions at $\sqrt{s}$ = 8 TeV with the ATLAS detector, Eur. Phys. J. C, 79, 760 (2019)

\bibitem{Aaij:2012vn}
R. Aaij et al., Inclusive W and Z production in the forward region at $\sqrt{s}$ = 7 TeV, JHEP, 06, 058 (2012)
\bibitem{Aaij:2014wba}
R. Aaij et al., Measurement of the forward W boson cross-section in pp collisions at $\sqrt{s}$ = 7 TeV, JHEP, 12, 079 (2014)
\bibitem{Aaij:2015zlq}
R. Aaij et al., Measurement of forward W and Z boson production in pp collisions at $\sqrt{s}$ = 8 TeV, JHEP, 01, 155 (2016)
\bibitem{Aaij:2016qqz}
R. Aaij et al., Measurement of forward $W\rightarrow e\nu$ production in pp collisions at $\sqrt{s}$ = 8 TeV, JHEP, 10, 030 (2016)

\bibitem{Li:2012wna}
Y. Li and F. Petriello, Combining QCD and electroweak corrections to dilepton production in FEWZ, Phys. Rev. D, 86, 094034 (2012)
\bibitem{Catani:2009sm}
S. Catani, L. Cieri, G. Ferrera, D. de Florian, M. Grazzini, Vector boson production at hadron colliders: a fully exclusive QCD calculation at NNLO, Phys. Rev. Lett., 103, 082001 (2009)

\bibitem{Grazzini:2017mhc}
M. Grazzini, S. Kallweit, and M. Wiesemann, Fully differential NNLO computations with MATRIX, Eur. Phys. J. C, 78, 537 (2018)
\bibitem{Catani:2007vq}
S. Catani and M. Grazzini, An NNLO subtraction formalism in hadron collisions and its application to Higgs boson production at the LHC, Phys. Rev. Lett., 98, 222002 (2007)
\bibitem{Catani:2012qa}
S. Catani, L. Cieri, D. de Florian, G. Ferrera, and M. Grazzini, Vector boson production at hadron colliders: hard-collinear coefficients at the NNLO, Eur. Phys. J. C, 72, 2195 (2012)

\bibitem{Matsuura:1988sm}
T. Matsuura, S. C. van der Marck, and W. L. van Neerven, The calculation of the second order soft and virtual contributions to the drell-yan cross-section, Nucl. Phys. B, 319, 570-622 (1989)
\bibitem{Cascioli:2011va}
F. Cascioli, P. Maierhofer, and S. Pozzorini, Scattering amplitudes with Open Loops, Phys. Rev. Lett., 108, 111601 (2012)
\bibitem{Denner:2016kdg}
A. Denner, S. Dittmaier, L. Hofer, Collier: a fortran-based complex one-loop lIbrary in extended regularizations, Comput. Phys. Commun., 212, 220-238 (2017)

\bibitem{Buckley:2014ana}
A. Buckley, J. Ferrando, S. Lloyd, K. Nordstrom, and B. Page, LHAPDF6: parton density access in the LHC precision era, Eur. Phys. J. C, 75, 132 (2015)
\bibitem{Ball:2014uwa}
R. D. Ball et al., Parton distributions for the LHC Run II, JHEP, 04, 040 (2015)
\bibitem{Dulat:2015mca}
S. Dulat, T. Hou, J. Gao, M. Guzzi, and J. Huston, New parton distribution functions from a global analysis of quantum chromodynamics, Phys. Rev. D, 93, 033006 (2016)
\bibitem{Harland-Lang:2014zoa}
L. A. Harland-Lang, A. D. Martin, P. Motylinski, and R. S. Thorne, Parton distributions in the LHC era: MMHT 2014 PDFs, Eur. Phys. J. C, 75, 204 (2015)

\bibitem{Butterworth:2015oua}
J. Butterworth et al., PDF4LHC recommendations for LHC Run II, J. Phys. G, 43, 023001 (2016)

\bibitem{Binoth:2000ps}
T. Binoth and G. Heinrich, An automatized algorithm to compute infrared divergent multiloop integrals, Nucl. Phys. B, 585, 741-759 (2000)

\end{thebibliography}
\end{document}